\documentstyle[aps,prl,epsf]{revtex}
\begin{document}

\twocolumn[\hsize\textwidth\columnwidth\hsize\csname
@twocolumnfalse\endcsname
\draft
\author{C.D. Batista, J.M. Eroles, M. Avignon\protect\cite{perm}
and B. Alascio}
\address{Centro At\'{o}mico Bariloche and Instituto Balseiro, (8400) S. C. de Bariloche, Argentina.}

\title{Electron-Doped Manganese Perovskites: The Polaronic State}
\maketitle
\begin{abstract}
Using the Lanczos method in linear chains we study the ground state of the
double exchange model including an antiferromagnetic super-exchange in the low 
concentration limit. We find that this ground state is always inhomogeneous, containig
ferromagnetic polarons. The extention of the polaron spin
distortion, the dispersion relation and their trapping by impurities, are studied for
diferent values of the super exchange interaction and magnetic field.
We also find repulsive polaron polaron interaction.

\end{abstract}

\vskip2pc]

The discovery of 'colossal ' magnetoresistance (CMR) \cite{helm} together
with its many unusual properties has received considerable attention lately.

Experiments have revealed very rich phase diagram interpreted in terms of
magnetic ferro, antiferro, canted, polaronic, and non-saturated phases
. Charge ordered phases also have been found\cite{Ramirez}. The
phase diagram, as a function of concentration $x$, temperature, magnetic
field, or magnitude of the superexchange interaction is not quite clear yet
for the different compounds. The metallic phase can be reached by hole
doping of the parent compound LaMnO$_{3}$, by substituting La for divalent
alkalies, Pb, or by stoichiometry changes. Very recently, neutron scattering
experiments have been interpreted in terms of polaronic droplets\cite
{Hennion}. Much less is known about the electron doped compounds where
doping does not seem to produce metallization.

From the theoretical point of view, the pioneering work of de Gennes \cite
{degen} proposed a canted phase to resolve the competition between the
ferromagnetic double interaction introduced by the presence of itinerant
holes and the superexchange interaction. Recently, several contributions to
this problem have been reported. Arovas and Guinea \cite{Arovas}, studied
this problem using a Schwinger boson formalism to obtain a phase diagram
showing several homogeneous phases and pointing out that phase separation
replaces the canted phase in a large region. Indeed phase separation appears
in several numerical treatments of the problem \cite{Dagotto1}. In other
analytical treatments, more adequate to treat local instabilities,
non-saturated local magnetization states have appeared at zero temperature%
\cite{caty}. M.Yu.Kagan {\it et. al.} \cite{Khomsky} have studied the stability of the
canted phases against the formation of large ferromagnetic 'droplets'
containing several particles and they conclude that the formation of
droplets is favored in the ground state. The variety of results obtained
from the different approaches points to the need of clarifying the picture
and testing the results.

In this work, for the first time, we find the low energy quasiparticles and
characterize their structure and dispersion relation in the low
concentration limit. These quasiparticles correspond to the electron
followed by a ferromagnetic local distortion (ferromagnetic polaron) in the
antiferromagnetic (AF) background. The dispersion relation is dominated by $%
k\rightarrow k+\pi $ scattering due to the presence of AF order. In order to
make a conection with transport properties, we also study the the tendency
to localization of these polarons in the presence of impurities and magnetic
field.

To render evident the nature of the ground state, we resort to the Lanczos
method, which is free from approximations. The Hamiltonian is simplified to
a single orbital per site, no lattice effects are considered, and we have to
reduce to one dimensional chains. However our results provide a simple
picture that, we presume, can be put to test of the dilute limit of electron
doped systems. In these systems, the limitations of the model Hamiltonian
may not be as stringent as in the hole doped systems for the following
reasons: the lattice structure is more symmetric so Jahn Teller distortions
should play a less important role, the large in-site Coulomb repulsion
inhibits double occupation so that it may be possible to describe the
physics by the use of a single effective orbital, and finally the
antiferromagnetic structure of two interpenetrating lattices can be properly
described in one dimension.

In order to describe the manganites we consider two degrees of freedom:
localized spins that represent the $t_{2g}$ electrons at the Mn sites, and
itinerant electrons that hop from $e_{2g}$ $Mn$ orbitals to nearest neighbor 
$e_{2g}$ orbitals. The model Hamiltonian includes exchange $(J)$ energies,
an antiferromagnetic interaction between localized spins $(K)$ and a hopping
term of strength $t$ which we will use as energy unit hereafter. It reads: 
\begin{eqnarray}
H &=&-J_{h}\sum_{i}{\bf S}_{i}\cdot {\bf \sigma }_{i}+K\sum_{<i,j>}{\bf S}%
_{i}\cdot {\bf S}_{j}+ \\
&&\sum_{<i,j>,\sigma }t_{ij}\left( c_{i\sigma }^{+}\cdot c_{j\sigma
}+h.c.\right) \,
\end{eqnarray}

\smallskip where $n_{i,\sigma }=${\bf \ }$c_{i\sigma }^{+}$ $c_{i\sigma }$ ,
and $c_{i\sigma }^{+},$ $c_{i\sigma }$ creates and destroys an itinerant
electron with spin $\sigma $ at site $i,$ respectively. {\bf \ }${\bf S}_{i}$
and ${\bf \sigma }_{i}$ are the localized and itinerant spin operators at
site $i$, respectively. In what follows we take large values of J$_{h}$
which prevents double occupation and makes the on-site Coulomb interaction $%
U $ irrelevant. This model has been studied numerically for finite
concentration in reference\cite{Moreo}, and in the absence of AF coupling ($%
K=0$) in reference \cite{Dagotto1}. In this paper we focus on the dilute
limit.

\begin{figure}
\narrowtext
\epsfxsize=3.5truein
\epsfysize=2.5truein
\vbox{\hskip 1.15truein \epsffile{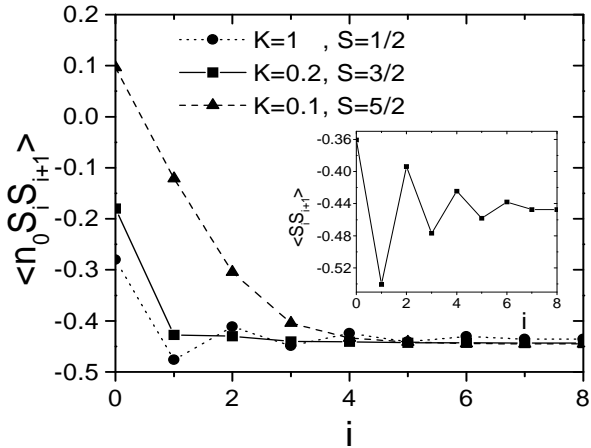}}
\medskip
\caption{We show the correlation function $<n_{0}S_{j}S_{j+1}>$ 
for a 16 sites chain, with $J_{h}=10$ and different values of $K$.
The maximun value of $S$ for each value of $K$ is indicated in the figure.
One can observe that the total $S$ and the extension of the magnetic
distorsion increase as $K$ decreases. The oscillatory behavior found at $K=1$
is a consequence of the weakening of the antiferromagnetic links around the
charge. The inset shows $<S_{j}S_{j+1}>$ for a 16 sites Heisenberg chain
where the link at site zero is a factor $2$ smaller than the rest.}
\end{figure}

We first investigate the homogeneity of the solutions for different values
of the antiferromagnetic interaction $K$ (in units of $t$). To this end, we
calculate the ground state with one electron added for chains of different
sizes up to $N=20$. With the aim of looking for spin distortions around the
charge, we calculate a correlation function which makes such a situation
evident : $<n_{i}S_{j}S_{j+1}>$. Because of translational symmetry the
results depend only on $\left| i-j\right| $. The results are shown in Fig. 1
where we plot $N<n_{0}S_{j}S_{j+1}>$ vs $j$, where $N$ is the number of
sites. As it can be seen in Fig. 1, for large $j$, this correlation function
takes a value very close to the one obtained from the Bethe ansatz solution
for the Heisenberg chain, $<S_{j}S_{j+1}>\cong 0.443$. The extension and the
magnitude of the spin distortion around the particle increases as K
decreases. The oscillations observed in the curve corresponding to $K=1$ are
also observed for larger values of $K$. They are a consequence of the
weakening of the antiferromagnetic links around the charge position which
produce a sort of local spin dimerization. To prove this point, we show in
the inset the nearest neighbors spin-spin correlation functions for a
Heisenberg chain of the same size where the link at site zero is a factor of
two smaller than the rest.

However as K decreases, it is difficult to find an adequate approximation to
describe the large polaronic distortion. In order to obtain the effective
mass of these polarons, we investigate the dispersion relation for charge
excitations. To this end we calculate the lowest energy state for different
values of the momentum $k=2\pi n/N$ within the subspace where the total spin
is that of the ground state. In Fig. 2 we show the dispersion relation
scaled to the thermodynamic limit for $K=3$, $J_{h}=100$; $K=1$, $J_{h}=10$;
and $K=0.3$, $J_{h}=10$.

\begin{figure}
\epsfxsize=3.9truein
\epsfysize=3.truein
\vbox{\hskip 0.05truein \epsffile{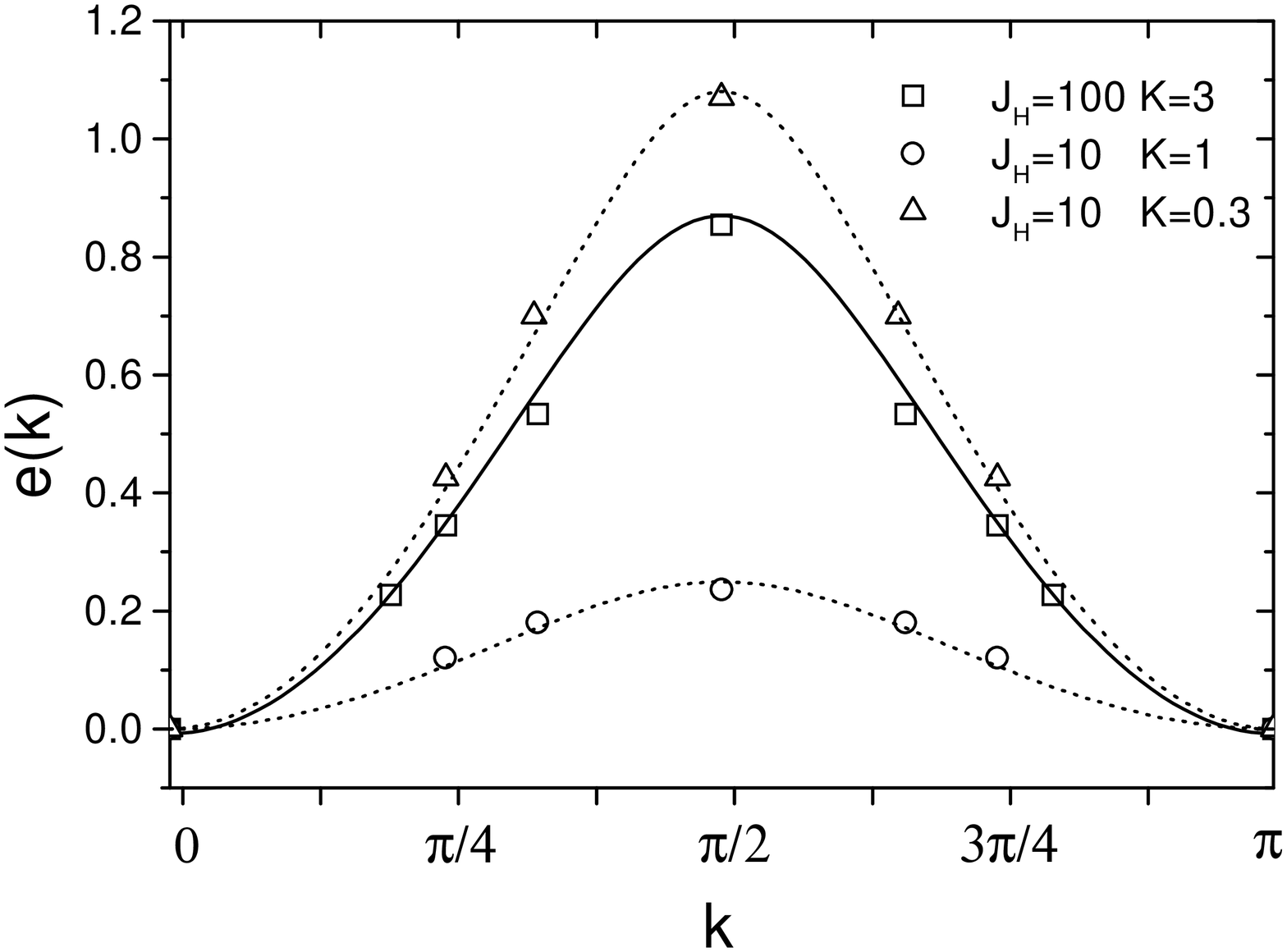}}
\caption{\label{fig:fig2} Dispersion relation scaled to the thermodynamic limit for
different sets of parameters. In full line we show a fit with the expression
discussed in the text for the limit $J_{H}>>k>t$ with $\Delta =1.4\approx K/2
$. In dotted line fit with $\Delta =1$ and $t=0.75$ for $K=1,$ and $\Delta
=0.6$ and $t=0.23$ for $K=0.3.$}
\end{figure}

We start analyzing the dynamics in the regime where $(J_{h}>>K>>t).$ In that
case, the charge moves as a spin one $({\bf \Sigma })$. The effective
hopping resulting from the projection of the hopping term onto the reduced $%
S=1$ Hilbert space is: $tP_{ij}({\bf \Sigma }_{i}{\bf S}_{j}+1/2)$, where $%
P_{ij}$ is the permutation operator between sites $i$ and $j$. We can
picture the movement of the particle, in this limit, as going from a state $%
\downarrow $ $\uparrow $ $\downarrow $ $\Uparrow $ $\downarrow $ $\uparrow $ 
$\downarrow $ to an intermediate state $\downarrow $ $\uparrow $ $\downarrow 
$ $\uparrow $ $0$ $\uparrow $ $\downarrow $, and finally to $\downarrow $ $%
\uparrow $ $\downarrow $ $\uparrow $ $\downarrow $ $\Uparrow $ $\downarrow ,$%
where $\Uparrow $ $(0)$ represents the $S_{z}=+1(0)$ components of the spin $%
S=1$. Thus, in order to move, the charge has to hop to the nearest neighbor,
via a spin flip process, through states that differ in energy by $\Delta
\approx K/2.$ It can be easily verified that the efective hopping of this
process is equal to $t_{ef}=t/\sqrt{2}.$ The dispersion relation given by
this dynamics is: $\Delta /2\pm \sqrt{(\Delta /2)^{2}+4t_{ef}^{2}\cos ^{2}(k)%
\text{ }}$ . The expression corresponding to the lower band is plotted with
full line in Fig. 2 and compared with the numerical result for $K=3$ and $%
J_{h}=100$. This expression is valid in general for a particle moving in an
antiferromagnetic background where scattering between $k$ and $k+\pi $
states dominates the dynamics of the particle (dotted lines in Fig.2).

In the case where $K>>J_{h}>>t$ the spin distortion can be neglected and the
particle propagates in an antiferromagnetic lattice. The Hund interaction
alternates the site energy of the propagating particle so that the
difference between the two sublattices is given by $\Delta =J_{h}(<\sigma
_{j}S_{j+1}>-<\sigma _{j}S_{j}>)\cong J_{h}(<S_{j}S_{j+1}>-<\sigma
_{j}S_{j}>)$ where we approximate $<\sigma _{j}S_{j}>\approx 1/4$ its value
at the triplet state, and $<\sigma _{j}S_{j+1}>\approx $ $<S_{j}S_{j+1}>=\ln
2-1/4$ , the Bethe ansatz value. Using these values we find $\Delta
=0.19J_{h}.$ In this case $t_{ef}$ is equal to $t.$

\begin{figure}
\epsfxsize=3.8truein
\epsfysize=3.truein
\vbox{\hskip 0.05truein \epsffile{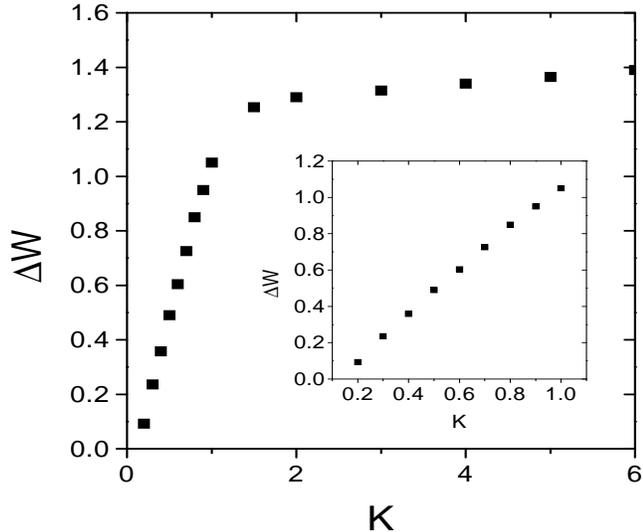}}
\caption{\label{fig:fig3} Effective bandwidth as a function of $K$. The change of regime as $%
K$ increases trough $t$ is evident.}
\end{figure}

When $J_{h}>>t\gtrsim K,$ the magnetic distortion around the charge is large
and the effective hopping is dominated by the overlap between the magnetic
distortions about the nearest neighbors sites. This last effect dominates
the polaron effective mass. Therefore, the mass of polarons increases when $K
$ decreases, as obtained in Fig. 2, where $t_{ef}$ decreases from 0.75 for $%
K=1$ to 0.23 for $K=0.3$, in agreement with the above results showing that
the spin distortion around the charge increases in magnitude and extension
when $K$ decreases. In Fig. 3, we calculate the bandwidth for several values
of $K$. One can distinguish clearly two regimes: $K<<t$ and $K>>t,$ the
first corresponds to a large magnetic distortion and the second corresponds
to a smaller one according to Fig. 1.

In order to test how robust is the polaronic description of the results, we
pin the polaron to a site by lowering in $\epsilon _{0}$ the diagonal energy
at site zero. This may be relevant to the real materials since the doping
process necessarily introduces some disorder. Since we are treating a linear
chain, this always localizes the particle, but the localization length
should be very different for different effective masses, so that a small $%
\epsilon _{0}$ localizes much more the polaron for low values of $K$ than
for larger ones. This is shown in Fig. 4 where we plot $<n_{i}>$ around site
zero for different values of $\epsilon _{0\text{ }}$and $K$. For comparison
we also show in full line the exponential fit of the different curves
showing the change in localization length.

\begin{figure}
\epsfxsize=3.5truein
\epsfysize=3.truein
\vbox{\hskip 0.05truein \epsffile{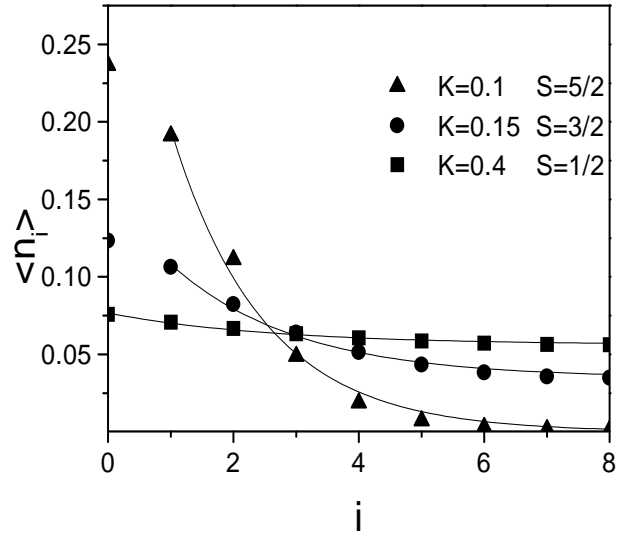}}
\caption{\label{fig:fig4}Figure 4. Charge localization. we show $<n_{i}>$ as a funcion of position in
a 16 sites chain where the energy at site zero is lowered from the rest by
0.05 units of $t$. Full lines are exponential fits to the curves with the
values of localization length $\lambda $ indicated in the figure. The short
localization length of the lower $K$ curves indicate effective bandwidths of
the order of the energy change.
}
\end{figure}

In Fig 5, we show the change in the values of $<n_{i}>$ for different
magnetic fields It can be seen that the localization of the polaron
decreases with magnetic field as a consequence of increased effective
hopping between nearest neighbors. A fact that may be important for the
transport properties of these systems since it implies a negative
magnetoresistive behavior for conductivity due to hopping between localized
states\cite{alalmex}

\begin{figure}
\epsfxsize=3.8truein
\epsfysize=2.5truein
\vbox{\hskip 0.05truein \epsffile{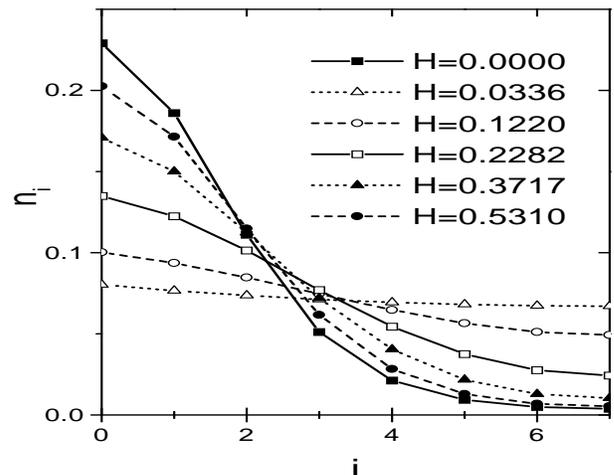}}
\caption{\label{fig:fig5}Figure 5. Effect of magnetic field on charge localization. We show $<n_{i}>$
for different values of an external magnetic field for the same chain as in
FIG. 4.}
\end{figure}

Finally, in Fig. 6 we present some of the results for the two particles
ground states. The charge-charge correlation function, normalized to the non
correlated $(J_{h}=0)$ case, clearly shows repulsion between the particles.
This long range repulsion increases with the magnetic distortion indicating
its magnetic origin.

\begin{figure}
\epsfxsize=3.5truein
\epsfysize=2.5truein
\vbox{\hskip 0.05truein \epsffile{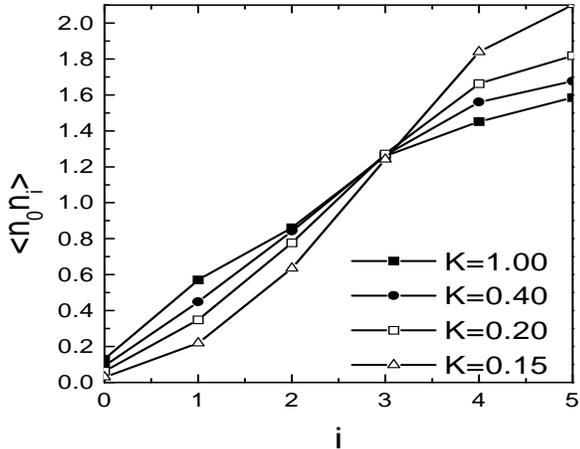}}
\caption{\label{fig:fig6} Figure 6. charge-charge correlation function for two particles in a ten site
chain for $J_{h}=10,$ and different values of $K.$ Values are normalized to
the uncorretated values.}
\end{figure}

The fact that polarons repel each other points to a picture of the electron
doped systems similar to that proposed originally by de Gennes \cite{degen}
of 'self-trapped electrons'. Further discussion on the non-diluted limit,
will be postponed for a later publication where newer results will be shown.

In summary, we have investigated the possibility of a non uniform ground
state in a model Hamiltonian using the Lanczos technique. The model
describes chains of localized spins coupled antiferromagnetically on which
electrons are added. These electrons suffer a strong ferromagnetic
interaction with the local spins and can hop from site to site. Assuming the
model adequately describes the physics of electron doped manganites, the
results presented here point to a picture of these systems where heavy one
electron polarons dominate the magnetic and transport properties. Their
masses depend strongly on the relation between the hopping energy and the AF
superexchange interaction. Clearly, the doping itself will localize the
polarons so that transport will result from hopping between pinned sites.
Negative magnetoresistance should appear as a consequence of the decrease of
the pinning energy with magnetic field\cite{alalmex}.

Adding two particles to the chains we find long range repulsion between
them. This long range repulsion could give raise to charge ordering.

Finally, we would like to point out that the order of oxygen vacancies in $%
CaMnO_{3-\delta }$ makes real the possibility of one dimensional electron
paths in these materials\cite{mate}. We hope that our results will stimulate
more experimental and theoretical investigations on the electron doped
manganites.

\begin{center}
{\it Acknowledgments}
\end{center}

Two of us (C.D.B. and J.M.E.) are supported by the Consejo Nacional de
Investigaciones Cient\'{i}ficas y T\'{e}cnicas (CONICET). B. A. is partially
supported by CONICET. M.A. is partially supported by the Centre National
pour la Recherche Scientifique\ (CNRS). We would like to aknowledge support
from the 'Fundacion Antorchas' and the Program ECOS-SECyT A97E05.

\end{document}